\documentclass[lettersize,journal]{IEEEtran}
\usepackage{cite}
\usepackage{amsmath,amsfonts}
\usepackage{algorithmic}
\usepackage{array}
\usepackage[caption=false,font=normalsize,labelfont=sf,textfont=sf]{subfig}
\usepackage{textcomp}
\usepackage{stfloats}
\usepackage{url}
\usepackage{verbatim}
\usepackage{graphicx}
\usepackage{algorithm}
\usepackage{algorithmic}

\usepackage{amsthm}
\usepackage{amssymb} 
\theoremstyle{definition}

\def\BibTeX{{\rm B\kern-.05em{\sc i\kern-.025em b}\kern-.08em
    T\kern-.1667em\lower.7ex\hbox{E}\kern-.125emX}}
\usepackage{balance}
\usepackage{listings}
\usepackage{xcolor}
\usepackage{booktabs}
\usepackage{colortbl}
\usepackage{multirow}

\usepackage{authblk}

\begin{document}

\title{Selecting New Measurement Locations to \\ Diversify Traffic-Pattern Coverage: A Real-World Evaluation for Total Traffic Volume Estimation}
\author[1]{Masaaki Inoue}
\author[2,3,4]{Akifumi Okuno}
\author[1]{Shintaro Fukushima}
\affil[1]{TOYOTA Motor Corporation}
\affil[2]{Institute of Statistical Mathematics}
\affil[3]{The Graduate University for Advanced Studies, SOKENDAI}
\affil[4]{RIKEN}

\markboth{Journal of \LaTeX\ Class Files,~Vol.~XX, No.~X, March~2026}%
{Shell \MakeLowercase{\textit{et al.}}: A Sample Article Using IEEEtran.cls for IEEE Journals}



\maketitle

\begin{abstract}
Accurate measurement of traffic volumes and flows is vital for modern intelligent transportation. 
However, despite recent technological advances in sensor devices, it is still expensive to install and maintain fixed traffic counters. 
Therefore, it is restricted to a small portion of location points where the counters can be installed, which severely limits the possibility of grasping and predicting the total traffic volume at a city-wide level. 
By contrast, devices with location history such as smartphones and connected vehicles are now widely used and provide much wider spatial coverage. However, the data from these devices are usually partial and noisy, so they are not enough to directly estimate total traffic volumes and flows. 
In this paper, we use the information from these widely available devices to help decide where to place additional traffic counters, and we study how selecting new measurement locations can improve city-wide traffic estimation performance.
To achieve this, we propose an algorithm that chooses additional counter locations to increase the diversity of observed traffic signal patterns, rather than simply spreading counters evenly over space. The goal is to capture traffic-pattern types that are rare in the current counter set and to make the collected observations more representative for later estimation and forecasting. 
We also present a real-world evaluation; in a target city, we select new locations expected to improve traffic prediction, and we then commissioned new field measurements at those locations at our expense. The resulting data led to an improvement in traffic volume estimation accuracy across different fidelities.
\end{abstract}


\section{Introduction}
\label{section:introduction} 

With recent technological advances in areas such as smart mobility, sensor devices, 
and information communication, it has become more and more important than ever to build intelligent transportation systems~\cite{ELASSY2024ITS}. 
In particular, it is one of the biggest issues in urban traffic management to understand and predict traffic flows~\cite{SHAYGAN2022review}. 
In this regard, traffic counters are often used for measuring traffic volume accurately at specific location points. 
However, it still requires a high cost for installing and maintaining the counters~\cite{SELBY2013kriging,Meng2017sigspatial}. 
Therefore, it is restricted to a small portion of location points where the counters can be installed. 
This fact severely limits the possibility of grasping and predicting the total traffic flow at a city-wide level; recently, portable devices with location history such as connected vehicles (CVs) and smart-phones have widely prevailed. 
With the increase of such devices, it is highly expected to estimate the total traffic flow at a city-wide level by leveraging data collected from the devices as well as the fixed traffic counters~\cite{Prokopec2013gps, HOTEIT2014mobile, Yong2022itsc}.

From an application point of view, it is also important to estimate the total traffic flow for the \textit{traffic digital twin}~(e.g., \cite{White2021Cities,Sánchez-Vaquerizo2021,Dasgupta2021arXiv,Lia02022TITS}) and road design, in addition to traffic management. 
To realize the traffic digital twin, it is necessary to collect traffic data in the real world, and then to reproduce the traffic flow in a digital~(virtual) world~\cite{xu2025enhancing}. 
However, a problem arises when we leverage the data collected from such devices; the ground truths~(the correct total traffic volume) are measured only at a limited amount of locations, and thus, the estimation accuracy for understanding total traffic flow is severely limited even when we combine the data from the portable devices~\cite{Owais2025sensor, Wu2023strategies}.

\begin{figure*}[!t]
\centering
\includegraphics[width=0.8\textwidth]{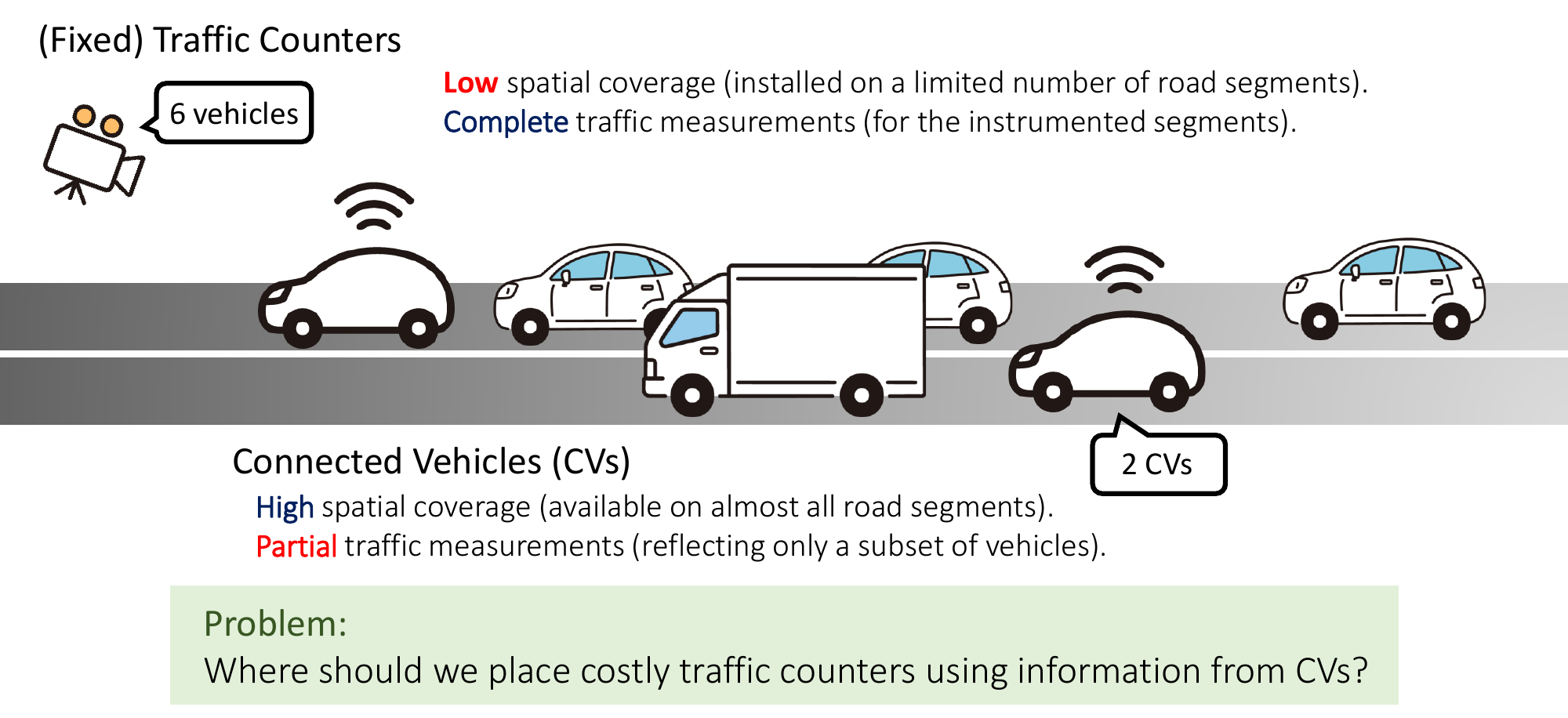}
\caption{Problem setting in this study. We combine two complementary data sources: 
\emph{Connected-vehicle (CV) data} is available on (almost) all roads, capturing high-coverage but partial traffic information. 
\emph{Fixed traffic counters} provide complete traffic volumes, but are costly, so they exist on only a small subset of road segments. 
Our goal is to use the (low-fidelity) broad CV view to decide where to deploy additional costly traffic counters.}
\label{fig:illustration}
\end{figure*}

Therefore, in this study, we address the issue of where we should add location points for measuring traffic to improve estimation accuracy, in addition to the location points of the fixed traffic counters. 
It should be noted that it is not possible to cover all the locations at a city-wide level under a realistic cost budget. 
For example, for the digital twin, we have to choose necessary and sufficient location points for reproducing the total traffic flow in a realistic way. 
This indicates that we should also prioritize important locations for understanding traffic flows. 
Therefore, our research question is stated as ``\textit{where should we additionally place traffic counters for understanding and predicting the total traffic flows under constraints of cost budget?}'' See Fig.~\ref{fig:illustration} for illustration.

To address this, we formulate the task of using widely available, low-cost location-history data (e.g., from connected vehicles and smartphones) to guide where to add a limited number of high-fidelity but costly sensors (i.e., traffic counters), with the goal of minimizing network-wide estimation error under realistic deployment constraints. 
Related work has explored geostatistical criteria such Euclidean distances~\cite{selby2013spatial} and road-network shortest-path distances~\cite{minjian2023location}; in contrast, our study determines sensor locations by leveraging similarity in road-level traffic patterns learned from low-cost portable devices such as CVs and smartphones. 
In our approach, we first characterize each road segment by traffic pattern features extracted from CV trajectories and embed all segments into a common feature space. 
We then estimate the distributions of segments with and without total traffic counters by kernel density estimation (KDE)~\cite{parzen1962estimation}.
Finally, we compute a KDE-based density ratio between the embedding distributions of segments with and without total traffic counters, and select segments with low ratios. A low ratio indicates traffic-pattern regions that are prevalent network-wide yet scarce among instrumented segments, suggesting segments whose pattern signals are not currently captured by existing counters. 

We establish a statistically sound evaluation protocol and show, on real city-scale datasets, that adding sensors selected by our method consistently improves downstream traffic-volume estimation compared with the original sensor configuration. Our contributions can be summarized in following points:
\begin{itemize}

\item We propose a site-selection algorithm that embeds road segments by waveform-shape similarity of standardized traffic patterns, estimates (via KDE) the embedding-space densities of (i) patterns observed on counter-equipped segments and (ii) patterns prevalent among segments observed only through portable devices, and then ranks candidate segments by a density-ratio score. Intuitively, a low ratio indicates segments whose traffic-pattern signals are common and thus important in the portable-device view of the network but are still underrepresented among counter-instrumented roads, so selecting such segments suggests where additional traffic counters should be deployed.

\item We establish an evaluation protocol that combines offline validation with a real-world prospective traffic survey: we deploy three-day-long measurements at locations selected by our method. We demonstrate on city-scale data that the additional measurements collected at the selected locations consistently improve downstream total-traffic estimation across different fidelities.
\end{itemize}

\section{Related Work}
\label{sec:related}

This section reviews prior studies from four perspectives: (i) network-wide traffic volume estimation,
(ii) sensor placement and observation site selection, (iii) representation learning for traffic patterns. 
Our key motivation is that the set of road segments equipped with ground-truth traffic volume counters often fails to represent the full diversity of network-wide traffic patterns. We therefore propose a sensor/site selection framework that explicitly targets and fills this representativeness gap in a learned traffic-pattern space, rather than optimizing a criterion tied to a particular downstream task.

\subsection{Traffic volume estimation }
\label{subsec:rw_volume}
A large body of work addresses network-wide traffic volume estimation by combining high-fidelity but spatially sparse fixed sensors (e.g., traffic counters) with broader-coverage probe sources, such as connected-vehicle (CV) trajectories, which are often biased or only partially observed~\cite{XING2024sparse}.
More recently, network-wide traffic prediction has increasingly relied on graph neural networks (GNNs) that directly encode road-network topology, including diffusion-based recurrent formulations and spatio-temporal graph convolutions~\cite{li2018dcrnn, ijcai2018GCN, Wu2019wavelet}.
While these methods can substantially improve prediction accuracy, their performance is often sensitive to the availability and spatial representativeness of ground-truth supervision~\cite{hein2025network}.
In practice, the limited number of instrumented road segments and the non-uniform placement of sensors can leave systematic gaps in coverage, constraining generalization to unobserved parts of the network~\cite{XING2024sparse}. This motivates research on where to collect additional ground-truth volumes to most effectively improve network-wide estimation.

\subsection{Sensor placement and observation site selection}
\label{subsec:rw_placement}
Sensor placement is commonly formulated as a budget-constrained optimization problem that aims to maximize information
or minimize estimation error~\cite{Joshi2009opt, Mehr2018submodular, LI2023submodular}. In optimal experimental design, locations are chosen to optimize criteria related to uncertainty
reduction or information gain such as~\cite{krause2008gaussian}. Complementary lines of
work adopt geometric or coverage-based objectives, such as~\cite{SALARI2019optlocation}.
A recurrent limitation, however, is that many criteria are coupled to a specific downstream model or error structure, and may require repeated retraining or uncertainty estimation to evaluate candidate sets. This model dependence can reduce portability across different
estimators and increases computational cost in large networks. 

\subsection{Representation and metric learning}
\label{subsec:rw_representation}
Representation learning for time series has advanced rapidly~\cite{franceschi2019unsupervised,eldele2021time}, enabling compact embeddings that capture temporal shape, periodicity, and
peak dynamics~\cite{zhang_self-supervised_2024}. Self-supervised methods (e.g., autoencoding and contrastive learning) are especially attractive in domains where labeled
ground truth is scarce, and have been used for clustering, retrieval, anomaly detection, and transfer learning~\cite{chen2020simple}. In traffic applications,
representation learning is often combined with graph-based modeling to incorporate network structure into node/link embeddings~\cite{yu2018stgcn,wu2019graph}.
In our setting, learned representations serve a distinct purpose: they provide a common latent space in which links can be compared based on
similarity of \emph{traffic patterns} even when ground-truth total volumes are missing. This enables the systematic assessment of which regions
of the traffic-pattern space are underrepresented by the currently instrumented segments, thereby turning sensor placement into a coverage problem over a learned manifold. 

\subsection{Summary and positioning}
\label{subsec:rw_summary}
Taken together, existing research provides tools for traffic volume estimation and principled sensor placement. Nevertheless, a gap remains between model-dependent placement criteria and
the practical need for scalable, transferable strategies that improve the coverage of traffic-pattern diversity under sparse supervision.
Our contribution is to connect traffic-pattern embeddings with density-ratio-based coverage scoring to select additional observation sites that
explicitly fill in the underrepresented regions of the traffic pattern space, offering a model-agnostic complement to conventional sensor placement approaches.

\section{Problem Formulation}
\label{sec:problem}

We formulate the problem of selecting additional measurement locations (e.g., fixed traffic counters including loop detectors, or manual traffic surveys) on a road network in order to improve network-wide estimation of total traffic volumes under a limited deployment budget.

\subsection{Road network and notation}
\label{sec:problem:network}

We consider an urban road network modeled as a directed graph
$G=(V,E)$, where $V$ is a set of nodes (e.g., intersections) and $E$ is a set of directed edges (e.g., road segments).
We discretize time into a finite set $\mathcal{T}=\{1,\dots,T\}$ (e.g., hourly intervals).

\vspace{0.5em} \noindent
\textbf{Total traffic volume:}
For each road segment $e\in E$ and time $t\in\mathcal{T}$, let $b_{t,e}\in\mathbb{R}_{\ge 0}$ denote the \emph{total traffic volume}, i.e., the number of vehicles passing through the road segment $e$ during time interval $t$. The total traffic volume is typically recorded by the costly traffic counters; they are observed only for a partial set of road segments. 

\vspace{0.5em} \noindent
\textbf{Partial traffic volume reported from CVs:} 
Similarly to the total traffic volume $b_{t,e}$, let $c_{t,e}\in\mathbb{R}_{\ge 0}$ denote the \emph{partial traffic volume} observed from connected vehicles (CVs) or location-history devices at $(t,e)$.
Intuitively, $c_{t,e}$ is a biased and incomplete proxy of $b_{t,e}$ because CV penetration and communication are not universal.

\vspace{0.5em} \noindent
\textbf{Optional static features:}
We optionally use time-invariant intersection and road segment attributes.
For each intersection $v \in V$, let $\mathbf{x}_v\in\mathbb{R}^{F_V}$ be its feature vector, and stack them as
$X_V=\{\mathbf{x}_v\}_{v\in V}$.
For each road segment $e\in E$, let $\mathbf{y}_e\in\mathbb{R}^{F_E}$ be its feature vector, and stack them as
$Y_E=\{\mathbf{y}_e\}_{e\in E}$.
When such features are unavailable, the framework reduces to using CV time series only.

\subsection{Measurement availability and feasible candidates}
\label{sec:problem:availability}

We assume that the partial traffic volumes reported from CVs $\{c_{t,e}\}_{t\in\mathcal{T}}$ are available for (almost) all road segments $e\in E$.
In contrast, ground-truth total traffic volumes $\{b_{t,e}\}_{t\in\mathcal{T}}$ are available only on a limited subset of road segments equipped with reliable sensors (or surveyed historically). Let
\begin{equation}
E_{\mathrm{available}} \subseteq E
\end{equation}
denote the set of road segments where total traffic volumes are already measured, i.e., $b_{t,e}$ is observed for all $t\in\mathcal{T}$ and $e \in E_{\mathrm{available}}$.

Because installing and maintaining sensors is costly and not all road segments are deployable, we define a \emph{feasible candidate set}
\begin{equation}
F \subseteq E \setminus E_{\mathrm{available}}
\end{equation}
that satisfies practical constraints such as site accessibility, safety, power/communications availability, and basic data-quality thresholds (e.g., minimum CV sample size or minimum road length).
Our goal is to select, from $F$, a limited number of road segments on which to deploy additional sensors so as to most effectively improve the traffic-volume prediction accuracy.

\subsection{Downstream estimation model and error}
\label{sec:problem:model}

Let $f(c, X_V, Y_E;\theta)$ be an estimation model parameterized by $\theta$ that predicts total traffic volumes from a partial count $c$ and available information $(X_V, Y_E)$.
For any $(t,e) \in \mathcal{T} \times E$, we form a fidelity-upscaling estimation
\begin{equation}
\hat{b}_{t,e} = f ( c_{t,e} , X_V, Y_E; \hat{\theta} ),
\label{eq:prediction_model}
\end{equation}
where $\hat{\theta}$ is an estimator.

Given a set $E_{\mathrm{available}}\subseteq E$ for which total counts are available, we define the training (or validation) loss by the squared error aggregated over time:
\begin{equation}
\mathcal{L}(E_{\mathrm{available}},\theta)
=
\sum_{t\in\mathcal{T}}
\sum_{e\in E_{\mathrm{available}}}
(\hat{b}_{t,e}-b_{t,e})^2.
\label{eq:empirical_loss}
\end{equation}

To emphasize that the sensor-placement strategy should not depend on a particular model form, we treat $f$ as a \emph{downstream} predictor: our selection method aims to improve prediction accuracy for a broad class of estimators trained on the augmented measurements.

\subsection{Additional measurement selection objective under budget}
\label{sec:problem:objective}

Let $E_{\mathrm{new}} \subseteq F$ denote the set of road segments selected for new measurements, and let $|E_{\mathrm{new}}|$ denote the number of selected segments.
We consider a \emph{budget} constraint on the number of additional sites:
\begin{equation}
|E_{\mathrm{new}}|\le B,
\label{eq:budget}
\end{equation}
where $B \in \mathbb{N}$ denotes the budget, specified by the planner. 
This simplified budget model is widely used in practice when per-site costs are difficult to quantify consistently at planning time.

To express the dependence of the trained model on the available measurements, let $\mathcal{A}$ denote a training procedure that maps a measurement set to fitted parameters:
\begin{equation}
\hat{\theta}(E_{\mathrm{available}}):=\mathcal{A}(E_{\mathrm{avalable}}).
\end{equation}
Our objective is to choose $E_{\mathrm{new}}$ so that the predictive error after augmentation is minimized.
Let $E_{\mathrm{test}}\subseteq E$ be the set of road segments used for evaluation. 
We formulate the sensor augmentation problem as finding $E_{\mathrm{new}} \subseteq F$ with the budget constraint~\eqref{eq:budget}, so that the new sensors reduce the prediction error:
\begin{align}
&\mathcal{L}\left(E_{\mathrm{test}}, \hat{\theta}(E_{\mathrm{available}}\cup E_{\mathrm{new}})\right) \nonumber \\
&\hspace{5em}\le \, 
\mathcal{L}\left(E_{\mathrm{test}}, \hat{\theta}(E_{\mathrm{available}})\right).
\label{eq:selection_problem}
\end{align}
This formulation captures the key planning question: \emph{which additional segments should be measured so as to improve network-wide prediction of total traffic volumes?} 
However, since solving this exactly is difficult, we treat the minimization problem as a working assumption and develop a heuristic yet systematic approach to selecting new measurement locations that is expected to reduce the prediction error.

\subsection{Offline evaluation protocol and assumptions}
\label{sec:problem:evaluation}

In real deployments, the total traffic volume $b_{t,e}$ is unknown on most road segments, so directly evaluating \eqref{eq:selection_problem} is impossible without installing sensors.
Therefore, we adopt an offline evaluation protocol using datasets where total counts are available on a sufficiently large subset of segments.

Let $E_{\mathrm{gt}}\subseteq E$ be the set of road segments for which ground-truth total traffic volumes are available in the dataset (i.e., segments instrumented with fixed traffic counters). 
We simulate the planning setting by splitting $E_{\mathrm{gt}}$ into:
(a) an existing measurement set $E_{\mathrm{available}}\subset E_{\mathrm{gt}}$,
(b) a candidate pool $F\subseteq E_{\mathrm{gt}}\setminus E_{\mathrm{available}}$,
and (c) a test set
\begin{equation}
E_{\mathrm{test}} \subseteq E_{\mathrm{gt}} \setminus (E_{\mathrm{available}}\cup E_{\mathrm{new}}).
\end{equation}
For each method, we:
(i) select $E_{\mathrm{new}}$ from $F$ under budget $B$ using only information available pre-deployment (e.g., CV volumes and static features),
(ii) reveal $\{b_{t,e}\}$ on $E_{\mathrm{new}}$ (simulating newly installed sensors),
(iii) train the downstream model on $E_{\mathrm{available}}\cup E_{\mathrm{new}}$,
and (iv) evaluate prediction error on $E_{\mathrm{test}}$ via the squared error~\eqref{eq:empirical_loss} (and report complementary metrics as needed).

We repeat the split (and, when relevant, random seeds for training) to obtain robust averages.
This protocol enables principled comparison of selection strategies while remaining faithful to the real constraint that total counts are initially sparse and expensive to obtain.

\section{Proposed Method}
\label{sec:method}

\begin{figure*}[t]
\centering 
\includegraphics[width=0.9\textwidth]{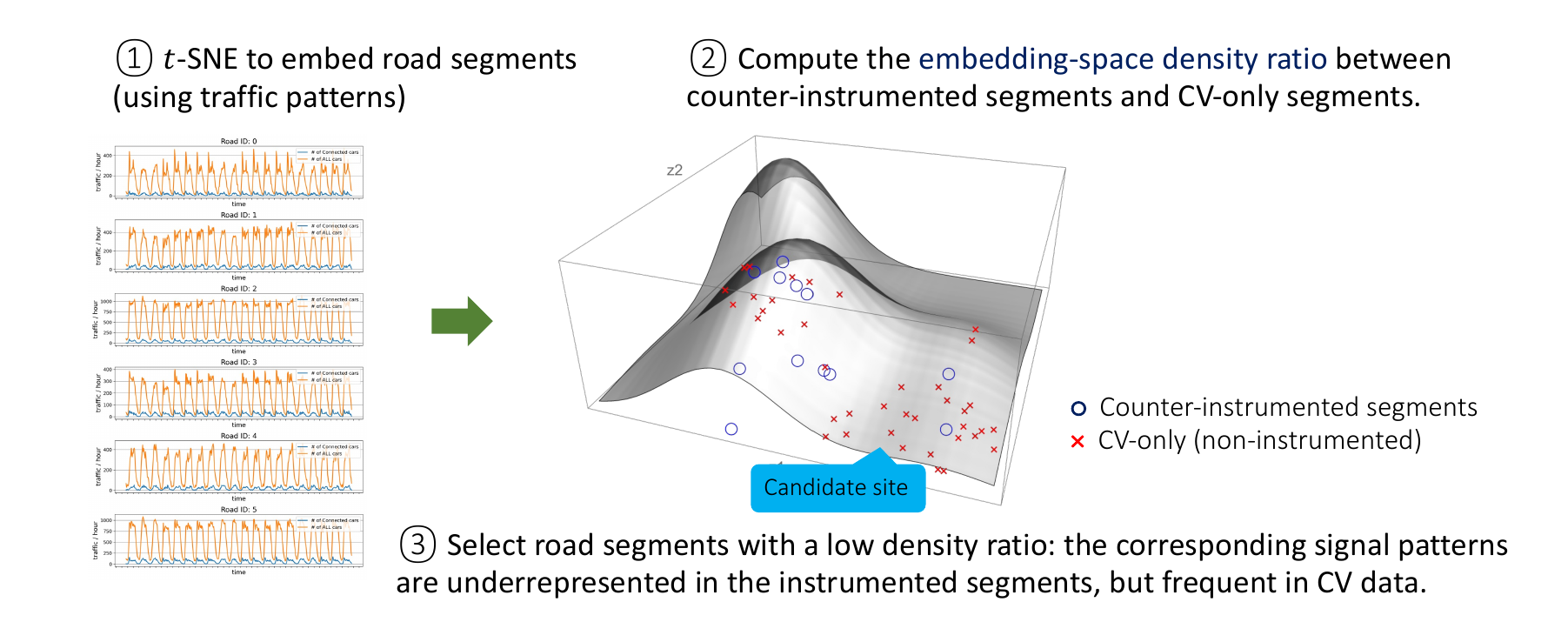}
\caption{Conceptual overview of our solution. Unlike ad hoc diversity heuristics (e.g., geographic gaps or road-category coverage), our approach promotes \emph{traffic-based diversity} by using waveform-shape similarity derived from CV traffic, which is available on any road. We first embed road segments to a Euclidean space via $t$-SNE, and select road segments whose signal patterns are frequent in CV data but underrepresented in the counter-instrumented segments.}
\end{figure*}

\subsection{Overview}
\label{subsec:overview}

Under a limited survey budget, this study aims to decide \emph{where to place costly traffic counters next} so as to most effectively improve a downstream estimator of total traffic.
A central challenge is that ground-truth supervision is spatially sparse, and the currently instrumented locations may fail to cover the full diversity of network-wide traffic dynamics.
Rather than relying on ad hoc notions of ``diversity'' (e.g., geometric spacing or road-class coverage), we frame the problem as \emph{coverage of traffic-pattern types} and explicitly target underrepresented regimes.

In our approach, we first use connected-vehicle (CV) traffic time series, which are available broadly across the road network, to learn a latent space in which road segments are organized by the \emph{waveform-shape similarity} of their traffic patterns. 
We then estimate, via kernel density estimation, the embedding-space density of currently instrumented segments and prioritize feasible candidates that lie in low-density regions (i.e., traffic-pattern regions that are rarely covered by the existing counter-instrumented segments). 
Concretely, the pipeline consists of: (i) embedding segment-level traffic patterns into a low-dimensional space, (ii) fitting KDE-based densities for the instrumented subset and for the full road network, and ranking candidates by a density-ratio score, and (iii) iteratively selecting road segments by referencing to the density ratio. 
The remainder of this section details each component: representation learning with wave-form similarity of traffic patterns (Section~\ref{subsec:sim}), density estimation and scoring (Section~\ref{subsec:dr}), and the iterative selection algorithm (Section~\ref{subsec:algorithm}).

\subsection{Embedding road segments via traffic patterns}
\label{subsec:sim}
\subsubsection{Waveform-shape similarity}
For each directed road segment $e\in E$, let $\mathbf{c}_e = (c_{1,e}, \dots, c_{T, e}) \in \mathbb{R}^{T}$ denote the CV traffic-volume time series (e.g., hourly counts).
To make heterogeneous segments comparable, we standardize each series to zero mean and unit variance:
\[
\begin{aligned}
c_{t,e}^{\mathrm{std}} &= \frac{c_{t,e}-\mu_e}{\sigma_e},\\
\mu_e &= \frac{1}{T}\sum_{t=1}^{T} c_{t,e},\\
\sigma_e^2 &= \frac{1}{T}\sum_{t=1}^{T} \bigl(c_{t,e}-\mu_e\bigr)^2,
\end{aligned}
\]
where $c_{t,e}$ denotes the observation at time $t$, and $c_{t,e}^{\mathrm{std}}$ is the standardized value. 
Let $\mathbf{c}^{\mathrm{std}}_e = (c_{1,e}^{\mathrm{std}}, \dots, c_{T,e}^{\mathrm{std}}) \in \mathbb{R}^T$ denote the standardized time series for segment $e$.

We then induce \emph{waveform-shape similarity} so that segments with similar temporal patterns are mapped nearby in a latent space.
Using the standardized series, we compute pairwise dissimilarities and convert them into affinities of the form
\begin{align}
s(e,e') \;\propto\;
\exp \left(-\dfrac{\|\mathbf{c}^{\mathrm{std}}_e-\mathbf{c}^{\mathrm{std}}_{e'}\|^2}{2 \tau^2} \right),
\label{eq:waveform-shape similarity}
\end{align}
where $\tau$ is a scale parameter (implicitly controlled in $t$-SNE via the perplexity setting).
We then obtain $t$-SNE coordinates that preserve local neighborhoods defined by these waveform-based similarities.

\subsubsection{Embedding road segments}
\label{subsubsec:t-SNE}
Using the waveform-shape similarity~\eqref{eq:waveform-shape similarity}, we compute $t$-distributed Stochastic Neighbor Embedding ($t$-SNE)~\cite{maaten2008tsne} to embed traffic patterns of road segments into a low-dimensional space. 
Particularly, each road segment $e$ is embedded as the Euclidean vector $\mathbf{z}_e \in \mathbb{R}^d$, where $d$ is the target embedding dimension.
We denote the embedding set over all segments by $Z=\{\mathbf{z}_e:\,e\in E\}$.

$t$-SNE is a nonlinear, neighborhood-preserving dimensionality reduction method that places samples with similar local structure close to each other in the embedding space.
Accordingly, the resulting coordinates are primarily intended to capture neighborhood relations rather than provide axes with direct physical interpretation.

\subsection{Density-ratio scoring}
\label{subsec:dr}
Let $E_{\mathrm{available}}\subseteq E$ denote the set of road segments for which the traffic volume is fully observed (i.e., ground-truth measurements are available), and let
\[
Z_{\mathrm{available}}=\{\mathbf{z}_e:\,e\in E_{\mathrm{available}}\},\,\,
Z_{\mathrm{all}}=\{\mathbf{z}_e:\,e\in E\}.
\]
We estimate two nonparametric densities in embedding space using Gaussian kernel density estimation (KDE)~\cite{parzen1962estimation,chen2017tutorial}:
a \emph{foreground} density $\hat p$ fitted on $Z_{\mathrm{available}}$ (i.e., density of the counter-instrumented segments) and a \emph{background} density $\hat q$ fitted on $Z_{\mathrm{all}}$ (i.e., density of the all segments):
\[
\begin{aligned}
\hat{p}(\mathbf{z})&=\frac{1}{|Z_{\mathrm{available}}|\,h_p^d}\sum_{\mathbf{u}\in Z_{\mathrm{available}}}K\!\left(\tfrac{\mathbf{z}-\mathbf{u}}{h_p}\right), \\
\hat{q}(\mathbf{z})&=\frac{1}{|Z_{\mathrm{all}}|\,h_q^d}\sum_{\mathbf{v}\in Z_{\mathrm{all}}}K\!\left(\tfrac{\mathbf{z}-\mathbf{v}}{h_q}\right),
\end{aligned}
\]
where $K(\cdot)$ is the standard Gaussian kernel and $h_p,h_q>0$ are bandwidths controlling smoothness.

\subsubsection{Bandwidth selection}
\label{subsec:bandwidth}
Bandwidths $(h_p,h_q)$ are chosen adaptively by cross-validation that maximizes held-out log-likelihood in the embedding space.
In the iterative procedure, $\hat p$ and thus $h_p$ is re-estimated after each addition to $E_{\mathrm{available}}$.
Since $Z$ is fixed, $\hat q$ (and $h_q$) can be selected once and cached.

\subsubsection{Density-ratio scoring}
For any $\mathbf{z}_e$ representing the embedded road segment $e$, 
we define the density-ratio score
\[
\hat{r}(e)=\frac{\hat{p}(\mathbf{z}_e)}{\hat{q}(\mathbf{z}_e)}.
\]
Intuitively, $\hat p$ and $\hat q$ are density estimates over the feature representation $\mathbf{z}$ that encodes each road segment's traffic-volume time-series pattern. In this sense, these densities quantify how common (or rare) a particular temporal pattern is within the corresponding segment set. Therefore, a small ratio $\hat{r}(e)=\hat{p}(\mathbf{z}_e)/\hat{q}(\mathbf{z}_e))$ indicates a pattern that is frequently observed in the CV-based data (high $\hat{q}$) but is missing or under-represented among the segments with fully observed, fixed-sensor measurements (low $\hat{p}$). Such segments are promising candidates for additional sensing because adding them can fill this representation gap and thereby improve prediction at other locations.

\subsubsection{Feasibility Constraints}
\label{subsec:constraints}
We restrict selection to a feasible set $\mathcal{F}\subseteq E$ to ensure practical deployability and data quality:
\[
\mathcal{F}=\{\,e\in E:\ \text{quality/feasibility constraints are satisfied}\,\}.
\]
Examples include minimum CV sample counts, installation constraints, exclusion zones, or diversification rules (e.g., avoiding overly close placements). 

\subsection{Iterative selection algorithm}
\label{subsec:algorithm}
\subsubsection{Procedure}
\label{subsec:stopping_outputs}
As shown in Algorithm~\ref{alg:method}, we iteratively update the foreground density and select the segment with the smallest density ratio. 
Starting from the current instrumented set $E_{\mathrm{available}}$, we iteratively add $B$ new sites.
At each step we refit the foreground density $\hat p$ on the updated $Z_{\mathrm{available}}$ and select the feasible uninstrumented segment with the smallest density ratio. The procedure stops when the budget cap is reached ($|E_{\mathrm{new}}|=B$) or when no feasible, uninstrumented candidates remain.
The output is an ordered list of selected segments, which can be used either (i) as a deployment plan or (ii) to evaluate downstream performance after augmentation.

\begin{algorithm}[t]
\caption{Candidate segment selection}
\label{alg:method}
\textbf{Input:} 
\begin{itemize} 
\item[] CV traffic patterns $\mathbf{c}_e$ for all segments $e\in E$; \item[] Counter-instrumented segment set $E_{\mathrm{available}}\subseteq E$;
\item[] Feasible segment set $F \subseteq E \setminus E_{\mathrm{available}}$;
\item[] Budget $B \in \mathbb{N}$.
\end{itemize}
\textbf{Output:} ordered list of candidate segments, that is denoted as \texttt{selected\_sites}.
\vspace{0.3em}\hrule\vspace{0.6em}

\begin{algorithmic}[1]

\STATE \textbf{Embed road segments:} Following Section~\ref{subsec:sim}, compute embeddings $\mathbf{z}_e\in\mathbb{R}^d$ for all road segments $e\in E$. 
\STATE \textbf{Compute background density:} Select the bandwidth $h_q$ by cross-validation on $Z$ and compute its KDE $\hat q$. 
\STATE Initialize $\texttt{selected\_sites}$ as an emptyset.

\FOR{$b=1$ \TO $B$}
    \STATE \textbf{Update foreground density:} Select the bandwidth $h_p$ by cross-validation on $Z_{\mathrm{available}}$ and compute KDE $\hat{p}$.
    \STATE \textbf{Select a candidate segment:} 
    \[
    e^\star \leftarrow \mathop{\arg\min}_{e \in F} \frac{\hat{p}(\mathbf{z}_e)}{\hat{q}(\mathbf{z}_e)}.
    \]
    \STATE \textbf{Update segment sets:} 
    \begin{itemize}
    \item[] Append $e^\star$ to $\texttt{selected\_sites}$;
    \item[] Remove $e^\star$ from $F$;
    \item[] Append $\mathbf{z}_{e^\star}$ to $Z_{\mathrm{available}}$.
    \end{itemize}
\ENDFOR

\end{algorithmic}
\end{algorithm}

\subsubsection{Complexity and Numerical Stability}
\label{subsec:complexity}
For $n=|E|$ segments and $d$-dimensional embeddings, naive KDE evaluation against all kernel centers is $O(n d)$ per query point; scoring all candidates is thus $O(n^2 d)$ per iteration in the worst case.
In practice, this can be accelerated using approximate to prune far-field contributions and by caching $\hat{q}$.

\section{Data \& Preprocessing}
\label{sec:data}

\subsection{Data sources}
\label{subsec:data_sources}
We use two complementary data sources to gather the traffic volumes and a road network of Toyota-city in Japan.
First, we use the open traffic-volume dataset provided by the Japan Road Traffic Information Center (JARTIC)~\cite{JARTIC} as the primary source of reliable cross-sectional traffic counts.
The JARTIC open data provides directional traffic volumes for instrumented road sections together with geospatial attributes (e.g., location/geometry identifiers), and can be downloaded from \cite{JARTIC}.
We use these measurements as the ground-truth total traffic volume \(b_{t,e}\) available only on a subset of road segments \(E_{\mathrm{available}}\subseteq E\).
Second, we use CV trajectory data and road network data obtained from a private TOYOTA database\footnote{This dataset is not publicly available due to privacy and data protection considerations.}. 
We map-match CV trajectories to the road network and aggregate them into hourly traffic volumes for each road segment, denoted by $c_{t,e}$. 

Our JARTIC and CV traffic data spans \(720\) hours (i.e. 30 days) from November 1, 2024, to November 30, 2024, yielding the partial-traffic time series $\mathbf{c}_e=(c_{1,e},\ldots,c_{T,e})^\top$ with $T=720$ for the road segments with sufficient coverage.
For privacy and confidentiality, we only use temporally aggregated, segment-level counts in the modeling pipeline.

\subsection{Road network and spatial linkage}
\label{subsec:spatial_linkage}
We represent the study area as a directed road network \(G=(V,E)\).
The road network geometry and topology (nodes, directed links, and their attributes) are obtained from a private TOYOTA road-network database. 

We spatially align all datasets to this common network.
Each JARTIC measurement record is linked to a directed road segment in $E$ using its geospatial attributes.
When a single measurement section overlaps multiple candidate segments (e.g., near complex junction geometries), we apply deterministic tie-breaking rules based on (i) geometric proximity, (ii) direction consistency, and (iii) maximum overlap length, so that each measurement record is mapped to a unique directed segment.
After linkage, the set of mapped segments defines $E_{\mathrm{available}}$, i.e., road segments whose traffic volumes $b_{t,e}$ are observed.
Similarly, CV trajectories are map-matched to segments in $E$ and aggregated into hourly counts to construct $c_{t,e}$ on a broad subset.

\subsection{Temporal alignment and aggregation}
\label{subsec:temporal_alignment}
The JARTIC open data are provided in fixed 5-minute aggregation intervals.
To fit the regular-interval time-series setting used in the remainder of this paper, we align all observations to a common hourly time grid $t=1,\ldots,T$ over the analysis period $\mathcal{T}$.
Since the original JARTIC data are 5-minute counts, we aggregate them to hourly counts by summation within each hour.
The CV trajectories are aggregated to the same hourly grid by counting (map-matched) CV traversals per directed road segment per hour.
Then the traffic volumes for CVs and all vehicles are assigned to a single value for each $(t,e)$, ensuring consistent indexing over $\mathcal{T}$.

\subsection{Feature construction and normalization}
\label{subsec:feature_construction}
For each directed road segment \(e\), we construct a partial-traffic time series
$\mathbf{c}_e=(c_{1,e},\ldots,c_{T,e})^\top\in\mathbb{R}^{T}$ sampled at regular hourly intervals over $\mathcal{T}$ with $T=720$ (i.e. 720 hours) in our main setting.
Prior to road-segment embedding, we standardize each segment's traffic patterns to remove scale differences across heterogeneous roads:
\begin{align}
&\mathbf{c}_{e}^{\mathrm{std}}=\frac{c_{t,e}-\mu_e}{\sigma_e},\nonumber \\
&\mu_e=\frac{1}{720}\sum_{t=1}^{720}c_{t,e},\quad
 \sigma_e^2=\frac{1}{720}\sum_{t=1}^{720}(c_{t,e}-\mu_e)^2. \nonumber
\end{align}
where $c_{t,e}$ denotes the $t$th temporal sample at the segment~$e$.
This per-segment normalization yields standardized inputs that emphasize waveform-shape similarity (e.g., peak timing and variability) rather than absolute magnitude.

\section{Traffic Surveys and Real-World Evaluation}
\label{subsec:feasible_set}
To prospectively evaluate the suggested sites under real measurements, we conducted independent traffic surveys at \(23\) locations selected from the feasible candidate set and a held-out validation set:
\(15\) locations are the proposed sites selected by our method, and the remaining \(8\) locations are reserved for validation.
\begin{figure}[t]
  \centering
  \includegraphics[width=\linewidth]{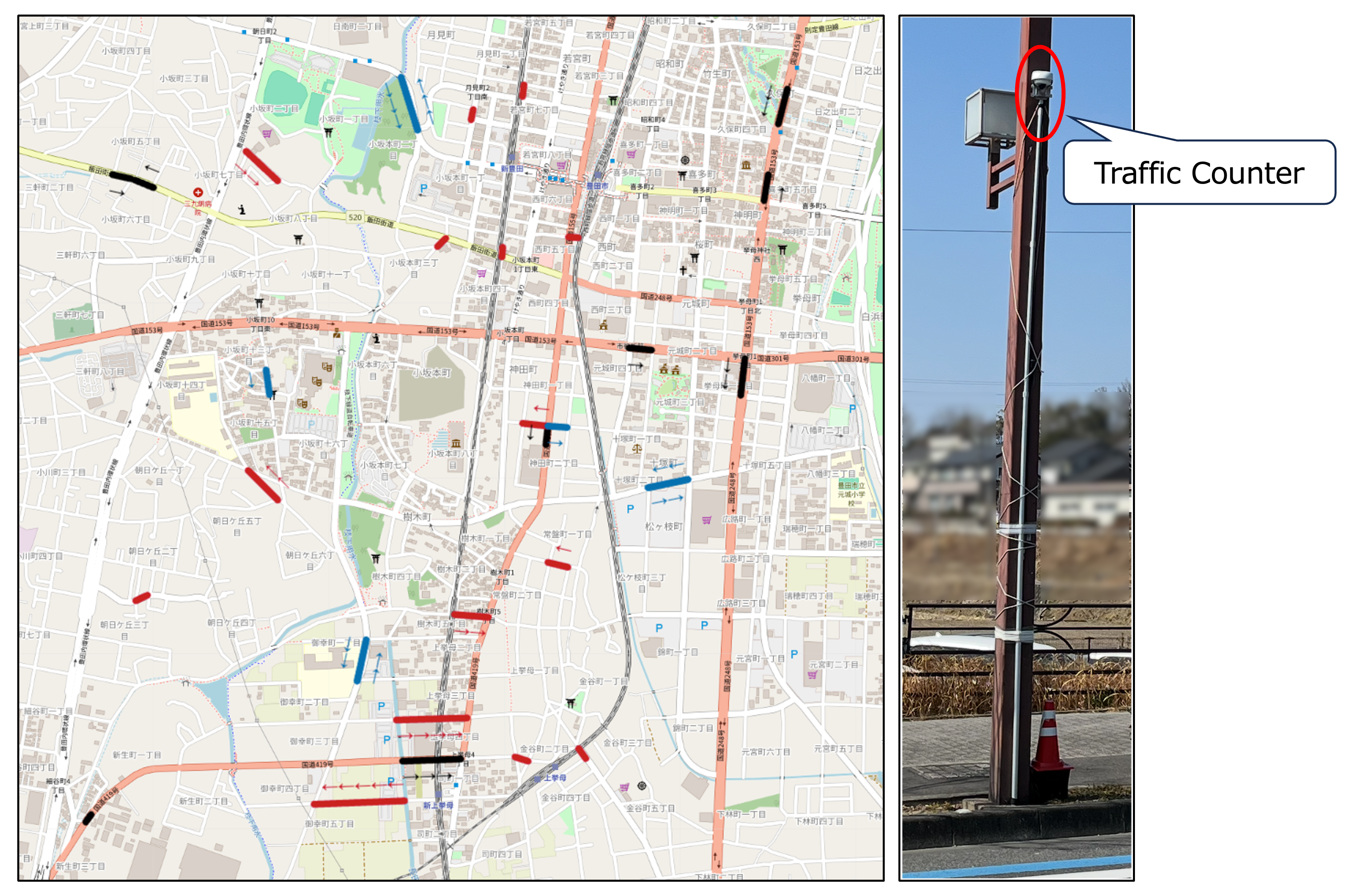}
    \caption{Left: Map of Toyota City showing the sites where JARTIC is available (black) and the newly surveyed locations (red and blue). Red segments denote the promising road segments selected by the proposed method, and blue segments denote the held-out road segments reserved for validation. Right: Example of traffic counter installation in the field.}
  \label{fig:map}
\end{figure}
Fig.~\ref{fig:map} visualizes the map of JARTIC measurement sites and the newly surveyed locations, where red segments indicate the promising road segments selected by the proposed method and blue segments indicate the held-out road segments reserved for validation. Maps were visualized using OpenStreetMap basemap tiles~\cite{openstreetmap} rendered with the Python \texttt{folium} library.

Traffic volumes were measured on February 24, 26, and 27, 2025, during 7:00--19:00.
These survey measurements provide additional ground-truth traffic counts for evaluation and are not used in constructing the embedding space.
Recall that not all road segments are deployable for the measurement planning. 
We therefore restrict selection to a feasible set \(F \subseteq E\) that satisfies installation and data-quality constraints.
In practice, \(F\) is constructed by considering the data quality (e.g., minimum CV coverage to ensure stable representation learning, minimum road segment length) and exclusion of segments already in \(E_{\mathrm{available}}\)).
We denote the resulting candidate set by
\begin{align*}
F=\{e\in E:\text{quality/feasibility constraints are satisfied}\}.
\end{align*}
All additional sites are selected from \(F\setminus E_{\mathrm{available}}\).

The field surveys were conducted after obtaining the necessary permissions from the relevant administrative authorities and were implemented with careful consideration of operational safety during sensor installation and measurement.
Furthermore, the surveys were designed to minimize privacy risks: the measurements targeted only traffic volumes at the aggregate level, and no personally identifiable information was acquired or retained.

\subsection{Feasible candidate set}
\label{subsec:feasible_set}

Not all road segments are suitable for deployment due to practical installation constraints and data-quality requirements. Accordingly, we restrict selection to a feasible candidate set $F \subseteq E$ that satisfies both deployability and quality criteria and excludes segments already contained in $E_{\mathrm{available}}$.

In our traffic survey, $F$ is constructed by imposing the following thresholds:
(i) a minimum connected-vehicle (CV) traffic volume of at least 1000 vehicles per month, and
(ii) a minimum segment length of at least 15 meters.
These constraints are introduced to avoid selecting segments that are too small-scale for reliable measurement, improving the stability of observed traffic volumes, satisfying practical requirements for traffic counting deployment, and reducing the risk of GPS mis-detections.
All site-selection experiments are conducted on $F$.

\subsection{Evaluation usage of survey measurements}
\label{subsec:survey_usage}

The newly obtained ground-truth counts at the 15 selected sites and the 8 held-out sites are used to evaluate whether the proposed method identifies practically informative segments.
Concretely, the surveyed counts enable quantifying the impact of augmenting $E_{\mathrm{available}}$ with newly measured segments on downstream, network-wide traffic-volume estimation. We emphasize that these newly surveyed traffics are used only for evaluation: they are not used to construct the embedding space nor to fit the density estimators, preserving the prospective nature of the assessment.

\subsection{Representation learning and temporal stability}
\label{subsec:repr_learning_results}

We embed each directed road segment using its standardized CV traffic-volume time series so that waveform-shape similarity is emphasized (e.g., peak timing and variability), rather than absolute traffic magnitude.
Specifically, for each road segment $e$ in the network, we construct as input a time series of hourly CAN-based traffic volumes,
\(
    \mathbf{y}_e(t) \in \mathbb{R}^{720},
\)
where the time index $t$ spans $720$ consecutive hours over a past month (from November 1, 2024, to November 30, 2024). 
Using these standardized segment-wise time series and their similarities defined in \eqref{eq:waveform-shape similarity}, we apply $t$-SNE to obtain a neighborhood-preserving embedding, and compress each segment representation into a $5$-dimensional vector $\mathbf{z}_e$ for subsequent density estimation.

To quantify temporal stability of similarity between waveform-shape of traffic between roads, we compare the distance structure among JARTIC sites across months.
For each selected month, we compute an $N \times N$ Euclidean distance matrix whose entries are the pairwise distances between standardized traffic time series at the $N$ JARTIC sites ($N=9$ in our experiment).
We then quantify the similarity between two months by taking the Pearson correlation coefficient between the off-diagonal elements of their distance matrices.
Repeating this for all month pairs yields the correlation matrix shown in Fig.~\ref{fig:time_invariance}.
The full mathematical definition of this distance-matrix correlation is provided in Appendix~A.

\begin{figure}[t]
  \centering
  \includegraphics[width=0.95\linewidth]{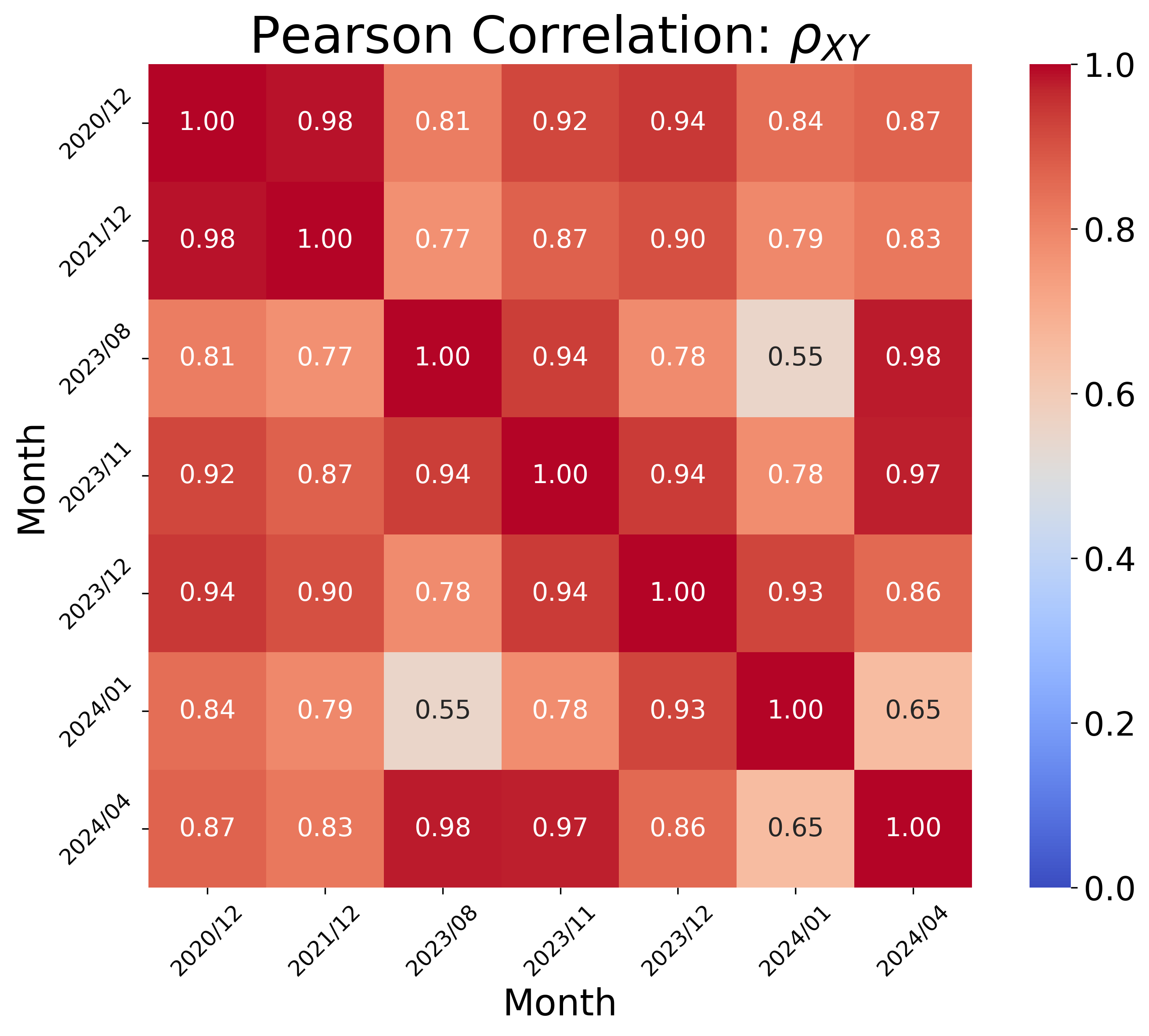}
  \caption{Pearson correlation matrix across months for the learned traffic-pattern representation (constructed from standardized CV traffic time series). High correlations indicate that the traffic-pattern structure is temporally stable and not dominated by a single snapshot.}
  \label{fig:time_invariance}
\end{figure}

\subsection{Density-ratio dynamics and selected sites}
\label{subsec:density_ratio_results}
Given the foreground density $\hat{p}$ estimated from currently instrumented segments and the background density $\hat{q}$ estimated from all segments, each feasible candidate segment $e \in F$ is scored by the density ratio $\hat{r}(e) = \hat{p}(\mathbf{z}_e)/\hat{q}(\mathbf{z}_e)$. As detailed in Algorithm~\ref{alg:method}, the proposed algorithm greedily selects segments whose density ratios are small.

\begin{figure*}[!t]
  \centering
  \includegraphics[width=\linewidth]{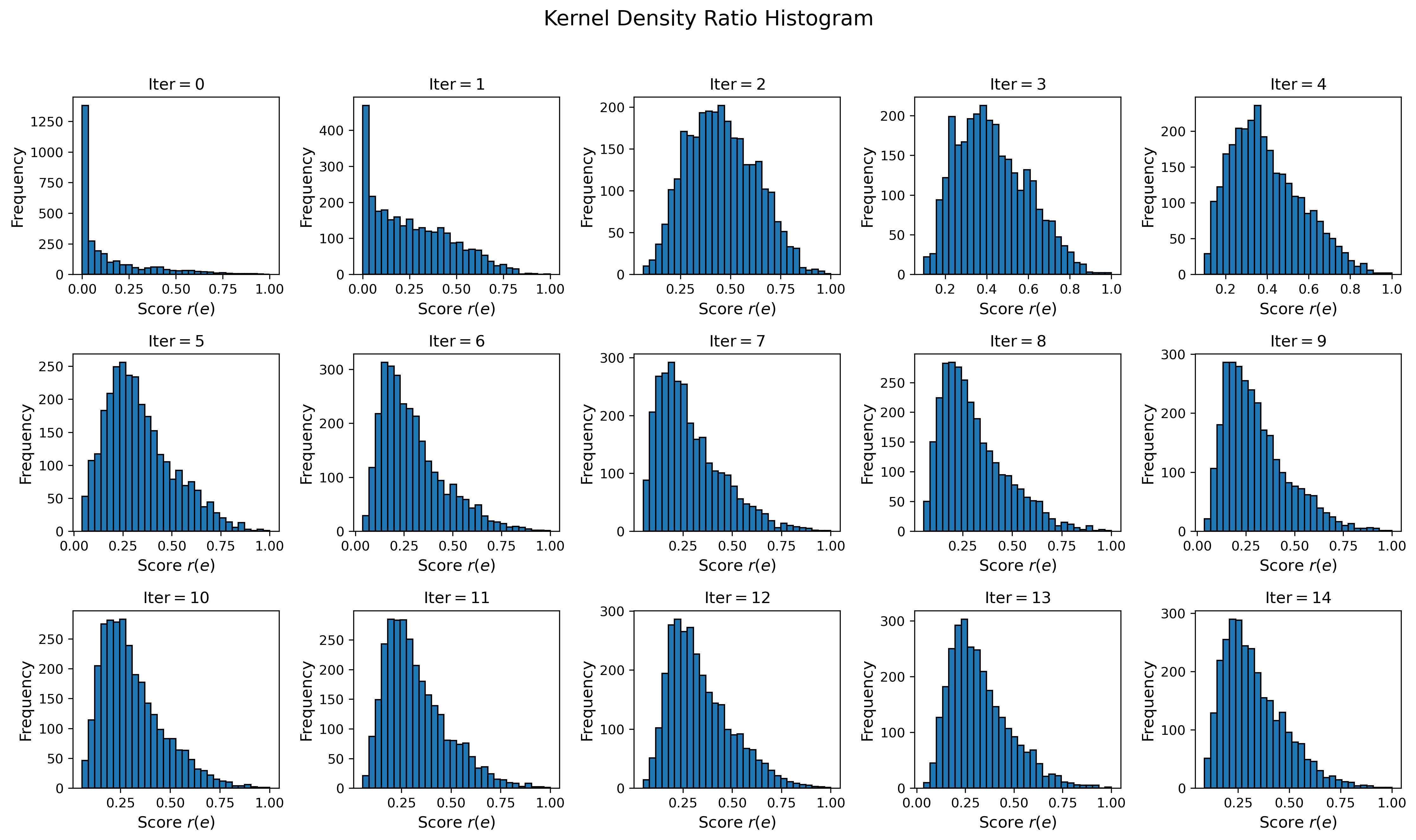}
  \caption{Distributional evolution of the density-ratio scores $r(e)=\hat{p}(\mathbf{z}_e)/\hat{q}(\mathbf{z}_e)$ over selection iterations. As additional sites are selected and added to the measured set, the score distribution shifts in a manner consistent with progressively filling representativeness gaps in embedding space.}
  \label{fig:kernel_density}
\end{figure*}

Fig.~\ref{fig:kernel_density} shows the distributional evolution of the density-ratio scores across iterations.
As additional sites are selected and appended to $E_{\text{exist}}$, the score distribution changes in a manner consistent with closing the representativeness gap: regions that were initially under-covered become progressively incorporated into the measured set.

\subsection{Impact on fidelity-upscaling total-traffic estimation}
\label{subsec:downstream_results}
We evaluate how augmenting ground-truth measurements using the proposed site-selection method affects downstream prediction of total traffic volume.
We consider following two prediction settings. The detailed formulations and implementation settings for both estimators are provided in Appendix~B.

\paragraph{Univariate regression (hour-wise scalar prediction)}
For each $t=(d,h)$ (where $d \in \{1,2,3\}$ and $h \in \{7,8,\ldots,18\}$ represent the and measurement day and hour-of-day, respectively), we predict the total traffic count $b_{t,e}$ from the CAN-based CV count observed at the same time: $\hat{b}_{t,e} = \hat{f}(c_{t,e})$, where $c_{t,e}\in\mathbb{R}$ denotes the CAN-based CV traffic volume at $t$ and $b_{t,e}\in\mathbb{R}$ denotes the corresponding ground-truth total traffic volume.
In this univariate setting, we consider two estimators: (i) a penetration-based method that scales the CAN-based CV count to total traffic using an assumed penetration rate, and (ii) univariate linear regression that learns a mapping from $c_{t,e}$ to $b_{t,e}$.

\paragraph{Multivariate regression (within-day waveform prediction)}
For each measurement day $d \in\{1,2,3\}$, we use 12 consecutive hours of observations and predict the corresponding 12-hour total-traffic waveform:
$\hat{\mathbf{b}}_{d,e} = \hat{\boldsymbol{f}}(\mathbf{c}_{d,e}) \in \mathbb{R}^{12}$, where $\mathbf{c}_{d,e} \in \mathbb{R}^{12}$ is the CAN-based CV traffic-volume waveform over the available 12-hour window (from 7{:}00 to 19{:}00) on day $d$ (day 1, 2, and 3), and $\mathbf{y}_{d,e}$ is the corresponding ground-truth total-traffic waveform over the same hours.
In this multivariate setting, we evaluate $\boldsymbol{f}$ using multivariate linear regression and two different architectures of neural networks (NN1: 1 hidden layer, 1000 units; NN2: 3 hidden layers, 100 units for each).

Table~\ref{tab:perform_comparison_uni} first summarizes the evaluation results for a univariate regression setting, comparing the post-deployment and original configuration of sensors in terms of MSE, MAE, and $R^2$ (for both the penetration-based method and univariate linear regression).
\begin{table}[t]
\centering
\caption{Performance comparison for univariate regression \\
(single time-point to single time-point)}
\label{tab:perform_comparison_uni}
\begin{tabular}{llrrr}
\toprule 
& & MSE & MAE & $R^2$ \\
\multirow{2}{*}{Penetration Method} &
\cellcolor[gray]{0.9}(Post-depl.) & \cellcolor[gray]{0.9} $6.6\times 10^{3}$ & \cellcolor[gray]{0.9} $5.7\times 10^{1}$ & \cellcolor[gray]{0.9} 0.60 \\
& (Original) & $6.8\times 10^{3}$ & $5.7\times 10^{1}$ & 0.58 \\
\multirow{2}{*}{Linear Regression} &
\cellcolor[gray]{0.9}(Post-depl.) & \cellcolor[gray]{0.9} $4.9 \times 10^{3}$ & \cellcolor[gray]{0.9} $5.1 \times 10^{1}$ & \cellcolor[gray]{0.9} 0.70 \\
& (Original) & $2.2\times 10^{4}$ & $1.4\times 10^{2}$ & -0.32 \\
\bottomrule
\end{tabular}
\end{table}
Table~\ref{tab:perform_comparison_multi} summarizes performance for three representative estimators: multivariate linear regression and two neural networks. 
Across all estimators, augmenting measurements using the proposed selection strategy yields lower MSE/MAE and higher $R^2$ than the original sensor configuration.
Notably, for multivariate linear regression, the proposed approach substantially reduces error and changes $R^2$ from negative (worse than predicting the mean) to a strong positive value, indicating that the additional measurements significantly improve generalization to unseen segments.

\begin{figure}[t]
  \centering
  \includegraphics[width=\linewidth]{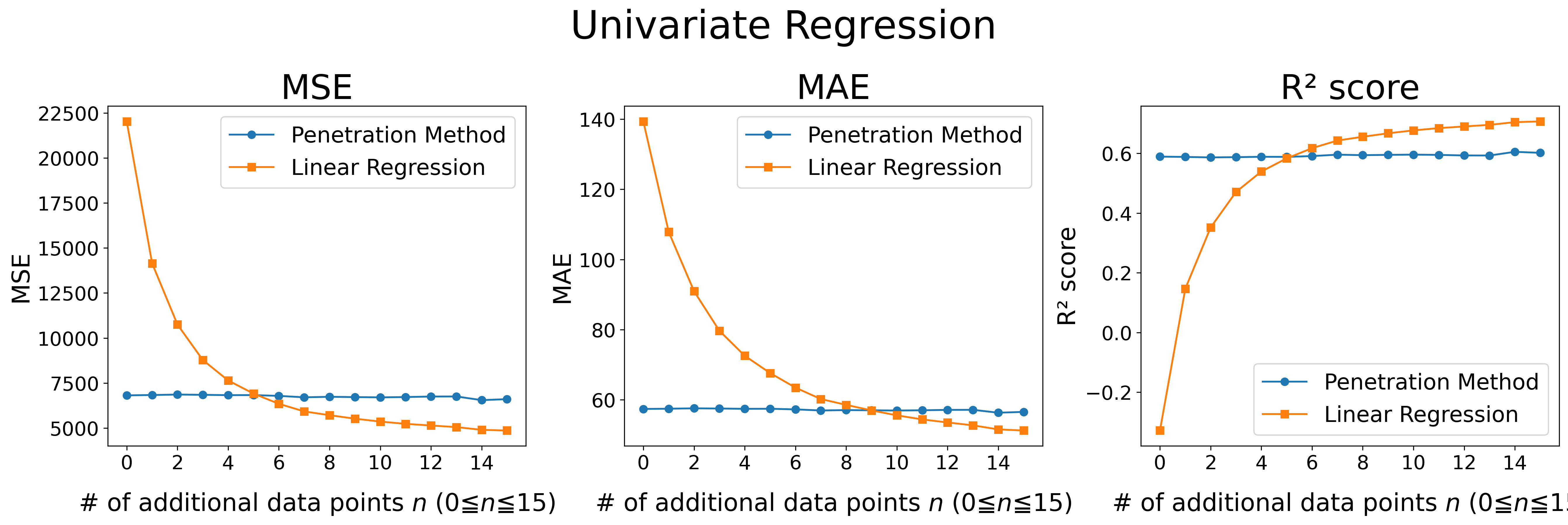}
  \caption{Univariate regression setting: each time point is mapped from low-fidelity CV data to high-fidelity measurements independently. Prediction performance as the number of additionally surveyed locations $n$ increases ($0\leq n \leq 15$). The three panels report MSE, MAE, and $R^2$ on held-out segments after augmenting the existing measurement set with the first $n$ selected sites.}
  \label{fig:uni_reg}
\end{figure}

\begin{figure}[t]
  \centering
  \includegraphics[width=\linewidth]{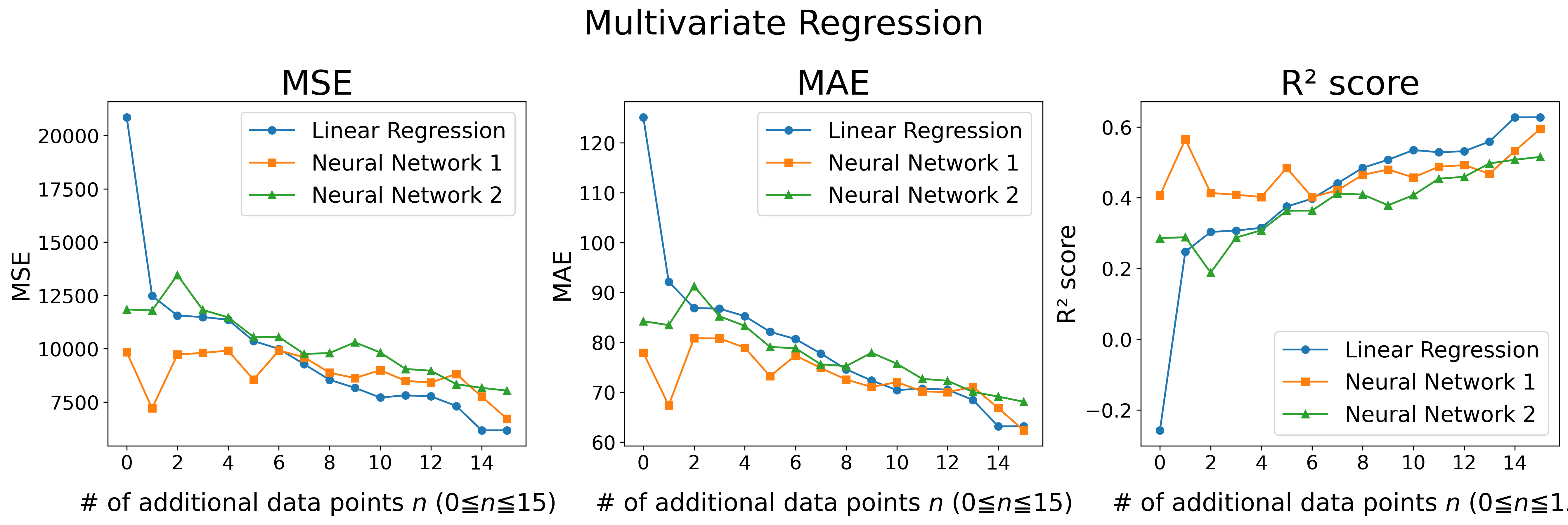}
  \caption{Multivariate regression setting: 12 time-points are mapped from low-fidelity CV data to high-fidelity measurements simultaneously. Prediction performance as the number of additionally surveyed locations $n$ increases ($0\leq n \leq 15$). The three panels report MSE, MAE, and $R^2$ on held-out segments for multivariate linear regression and two types of neural networks (NN1: 1 hidden layer, 1000 units; NN2: 3 hidden layers, 100 units for each).}
  \label{fig:multi_reg}
\end{figure}

\begin{table}[t]
\centering
\caption{Performance comparison for multivariate regression \\ 
(12 time-points to 12 time-points)}
\label{tab:perform_comparison_multi}
\begin{tabular}{llrrr}
\toprule
& & MSE & MAE & $R^2$ \\
\multirow{2}{*}{Linear Regression} &
\cellcolor[gray]{0.9}(Post-depl.) & \cellcolor[gray]{0.9} $6.2\times10^{3}$ & \cellcolor[gray]{0.9} $6.3\times10^{1}$ & \cellcolor[gray]{0.9} 0.62 \\
& (Original) & $2.1\times10^{4}$ & $1.3\times10^{2}$ & -0.25 \\
\multirow{2}{*}{Neural Network 1} &
\cellcolor[gray]{0.9}(Post-depl.) & \cellcolor[gray]{0.9} $6.8\times10^{3}$ & \cellcolor[gray]{0.9} $6.3\times10^{1}$ & \cellcolor[gray]{0.9} 0.58 \\
& (Original) & $8.1\times10^{3}$ & $7.0\times10^{1}$ & 0.51 \\
\multirow{2}{*}{Neural Network 2} &
\cellcolor[gray]{0.9}(Post-depl.) & \cellcolor[gray]{0.9} $5.7\times10^{3}$ & \cellcolor[gray]{0.9} $5.7\times10^{1}$ & \cellcolor[gray]{0.9} 0.65 \\
& (Original) & $1.1\times10^{4}$ & $8.6\times10^{1}$ & 0.31 \\
\bottomrule
\end{tabular}
\end{table}
Fig.~\ref{fig:uni_reg} further illustrates how prediction error evolves as the number of additional data points increases shown for a univariate regression setting.
The curves indicate that adding informative measurement sites leads to a monotonic reduction in MSE/MAE and an improvement in $R^2$, supporting the practical benefit of prioritizing sites that fill embedding-space representativeness gaps.
In addition, Fig.~\ref{fig:multi_reg} is shown for a \emph{multivariate} regression setting, and the same qualitative tendency is observed across all three predictors (linear regression and the two neural networks).
While the improvements are not uniform at every increment of $n$, the overall trajectories indicate decreasing MSE/MAE and increasing $R^2$ as additional data points are incorporated, suggesting that expanding measurements with informative sites yields progressively better predictive performance.

\subsection{Discussion}
\label{subsec:discussion_results}
Overall, the empirical results support the central premise of the proposed framework: the currently instrumented road segments do not necessarily cover the full diversity of network-wide traffic-pattern dynamics, and explicitly filling the coverage gaps in a learned traffic-pattern space improves downstream estimation. Because the selection criterion is defined via embedding-space densities rather than estimator-specific uncertainty, the approach is model-agnostic and yields consistent gains across heterogeneous predictors under realistic feasibility constraints.

A key implication is that the selection problem can be approached as a representativeness issue: the KDE-based foreground density (instrumented road segments) and background density (all segments) reveal where the measurement network under-covers common traffic-pattern regimes, and the density-ratio criterion systematically targets such regimes. The iterative evolution of density-ratio score distributions is consistent with progressively closing this representativeness gap as new sites are added.

From an application standpoint, improvements are observed not only for multivariate predictors but also univariate predictors. In particular, the proposed augmentation improves error metrics and $R^2$ in both Table~\ref{tab:perform_comparison_uni} (univariate regression) and Table~\ref{tab:perform_comparison_multi} (multivariate regression), and substantially changes cases where original sensor configuration is worse than mean prediction (negative $R^2$) into positive $R^2$ values. These results suggest that the benefit of measurement augmentation is not tied to a particular estimator class, but rather to improving the coverage of supervision over diverse traffic-pattern types.

The temporal stability of embedding also indicate that the learned traffic-pattern structure is not dominated by a single temporal snapshot: positive correlations across months support the use of embedding-space density estimates as a stable proxy for representativeness gaps, at least over the observation period considered.

There are several practical considerations and limitations.
Firstly, in the current formulation each road segment is treated with equal importance when computing background density. In practical deployments, however, the relative importance of segments can be heterogeneous. For example, when the prediction loss is non-uniform across the network because certain corridors, arterial roads, or critical segments are prioritized. In such cases, an important extension is to introduce segment-dependent weights so that the density-ratio scoring and the resulting site selection emphasize regions of the network that matter more for the downstream task.
Secondly, we rely on $t$-SNE as the sole dimensionality-reduction step to obtain compact representations. Since $t$-SNE is a nonlinear, neighborhood-preserving embedding, its coordinates are primarily meaningful in terms of local neighborhood relations and are not directly interpretable along individual axes; moreover, the resulting geometry can depend on hyperparameters (e.g., perplexity) and random initialization. Investigating alternative representation learning methods and density estimators that yield more interpretable embeddings is a promising direction.
Finally, although the density-ratio strategy is computationally scalable via caching the background density, further accelerations and extensions (e.g., graph simplification or alternative spatial discretizations) are important for large-scale deployment.

\section{Conclusion}
We studied the problem of determining additional traffic-measurement locations under a limited budget, motivated by the practical sparsity and non-uniform placement of ground-truth sensors in city-scale road networks.
Our proposed framework embeds road segments based on waveform-shape similarity of standardized CV traffic patterns, estimates embedding-space densities for instrumented segments and for all segments, and prioritizes new measurement sites using a KDE-based density-ratio score; low density-ratio indicates a pattern that is frequently observed in CV traffic time series, but is underrepresented among currently-instrumented segments. 

We evaluated the approach using an offline protocol and prospective traffic surveys, collecting new ground-truth counts at locations prioritized by our ranking and at held-out locations, and using them solely for evaluation. Augmenting the original measurements with these additional counts consistently improved downstream total-traffic estimation across multiple estimators and settings, reducing MSE/MAE and increasing $R^2$ relative to the original (no-addition) setting.

\section*{Acknowledgments}
This work is supported by TOYOTA Motor Corporation, and JSPS KAKENHI Grant Numbers 21K17718 and 25K03087.

\appendix[]
\section*{Appendix. A: Evaluation of temporal stability}
\label{app:distmat_corr}
Let $\mathcal{S}=\{s_1,\ldots,s_N\}$ denote the set of instrumented JARTIC sites (with $N=9$), and let $\mathbf{c}^{(m)}(s_i)\in\mathbb{R}^{T}$ denote the standardized traffic time series at site $s_i$ for month $m$.
For each month $m$, we define the pairwise Euclidean distance matrix $D^{(m)} = (d^{(m)}_{ij}) \in \mathbb{R}^{N\times N}$ by
\begin{equation}
d^{(m)}_{ij} = \left\| \mathbf{c}^{(m)}(s_i)-\mathbf{c}^{(m)}(s_j)\right\|_2,
\quad 1\le i,j\le N.
\end{equation}
We vectorize the unique off-diagonal entries (i.e., all unordered pairs) as
\begin{equation}
\mathbf{d}^{(m)} = ( d^{(m)}_{ij}: 1\le i<j\le N )\in\mathbb{R}^{N(N-1)/2}.
\end{equation}
Given two months $m$ and $m'$, we quantify the similarity between their distance structures by the Pearson correlation coefficient
\begin{equation}
\rho(m,m') \;=\; \mathrm{corr}\!\left(\mathbf{d}^{(m)},\mathbf{d}^{(m')}\right),
\end{equation}
where $\mathrm{corr}(\cdot,\cdot)$ denotes the standard Pearson correlation.
Computing $\rho(m,m')$ for all month pairs produces the month-to-month correlation matrix reported in Fig.~\ref{fig:time_invariance}.

\section*{Appendix. B: Evaluation Details}
\label{app:eval_details}

This section provides the details of prediction-task and methods. The main results are reported in Section~\ref{subsec:downstream_results}.

\subsection{Training/Test supervision sets (proposed vs.\ baseline)}
\label{app:supervision_sets}

CAN-based CV traffic volumes $\mathbf{c}_e$ are available for any segment $e \in E$. In contrast, ground-truth \emph{total} traffic volumes are available only at a limited number of locations. Therefore, supervised predictors are trained only on locations where both CAN-based CV volumes and total traffic counts are observed.

We compare two training-supervision configurations:
\begin{itemize}
\item \textbf{Proposal (augmented supervision):} training uses paired CAN-based CV volumes and total traffic volumes observed at the \emph{traffic survey locations} plus the existing \emph{JARTIC locations} (15 + 9 = 24 locations).
\item \textbf{Baseline (existing supervision only):} training uses paired CAN-based CV volumes and total traffic volumes observed at the \emph{JARTIC locations} only (9 locations).
\end{itemize}
In both configurations, CAN-based CV volumes can be obtained for any segment at inference time; the only difference is the amount and coverage of supervised training pairs.

\paragraph*{Test set.}
All models are evaluated on the same held-out \emph{validation locations} (8 locations), where both CAN-based CV volumes and total traffic volumes are available. Performance is computed by predicting total traffic at these 8 locations from their CAN-based CV volumes.

\subsection{Model details}
\label{app:model_details}

\subsubsection{Penetration method}
\label{app:penetration_method}

The penetration method assumes that total traffic is proportional to CAN-based CV traffic with a global penetration rate. We estimate a single scaling factor from the training data and then apply it to the test data.

Let $\mathcal{D}_{\mathrm{tr}}$ denote the set of supervised training observations (over the training locations and time indices used in the main evaluation setting). We estimate an effective penetration rate as the aggregated ratio
\begin{equation}
\hat{p} \;=\; \frac{\sum_{(t,e)\in\mathcal{D}_{\mathrm{tr}}} c_{t,e}}{\sum_{(t,e)\in\mathcal{D}_{\mathrm{tr}}} b_{t,e}},
\end{equation}
and predict total traffic by
\begin{equation}
\hat{b}_{t,e} \;=\; \frac{1}{\hat{p}}\,c_{t,e}.
\end{equation}
This method is used as an interpretable baseline in the univariate setting.

\subsubsection{Linear regression baselines}
\label{app:linear_regression}

\paragraph{Univariate linear regression.}
We fit an affine mapping from CAN-based CV volume to total volume:
\begin{equation}
\hat{b}_{t,e} \;=\; u \,c_{t,e} + \beta,
\end{equation}
where $(u,\beta)$ are learned from the training supervision set by least squares.

\paragraph{Multivariate linear regression.}
In the 12-hour waveform setting, we learn an affine mapping from $\mathbb{R}^{12}$ to $\mathbb{R}^{12}$:
\begin{equation}
\hat{\mathbf{b}}_{d} \;=\; \mathbf{U}\mathbf{c}_{d} + \boldsymbol{\beta},
\end{equation}
where $\mathbf{U}\in\mathbb{R}^{12\times 12}$ and $\boldsymbol{\beta}\in\mathbb{R}^{12}$ are learned by least squares.

\subsubsection{Neural network architectures}
\label{app:nn_architectures}

For the 12-hour waveform prediction, we evaluate two feed-forward neural networks with input dimension 12 and output dimension 12, using ReLU activations in hidden layers and a linear output layer:
\begin{itemize}
\item \textbf{Neural Network 1:} one hidden layer with 1000 units.
\item \textbf{Neural Network 2:} three hidden layers with 100 units each.
\end{itemize}

The corresponding Keras-style definitions are:
\begin{lstlisting}
# Neural Network 1
model = Sequential()
model.add(Dense(1000, input_dim=12, activation='relu'))
model.add(Dense(12, activation='linear'))

# Neural Network 2
model = Sequential()
model.add(Dense(100, input_dim=12, activation='relu'))
model.add(Dense(100, activation='relu'))
model.add(Dense(100, activation='relu'))
model.add(Dense(12, activation='linear'))
\end{lstlisting}


\section{References}

\bibliographystyle{IEEEtran}
\bibliography{main}

\newpage




\vfill

\end{document}